\pdfoutput=1
\documentclass[useAMS]{fpaletou}

\usepackage{graphicx}

\title[Inversion of stellar parameters from high-resolution Echelle
spectra]{Inversion of stellar parameters from high-resolution Echelle spectra}
\author[F. Paletou, J.-F. Trouilhet and
T. B\"{o}hm]{F. Paletou$^{1,2}$\thanks{E-mail: fpaletou@irap.omp.eu},
  J.-F. Trouilhet$^{1,2}$ \thanks{E-mail: jtrouilhet@irap.omp.eu} and T. B\"{o}hm
  $^{2,1}$\thanks{E-mail: tboehm@irap.omp.eu}\\
$^{1}$Universit\'e de Toulouse, UPS-Observatoire Midi-Pyr\'en\'ees,
Irap, Toulouse, France\\
$^{2}$ CNRS, Institut de Recherches en Astrophysique et
Plan\'etologie, 14 ave. E. Belin, F--31400 Toulouse, France}
\begin{document}

\date{Please send us any comments etc..}

\pagerange{\pageref{firstpage}--\pageref{lastpage}} \pubyear{2014}

\maketitle

\label{firstpage}

\begin{abstract}

The general context of this study is the inversion of stellar
fundamental parameters from high-resolution Echelle spectra.  We aim
indeed at developing a fast and reliable tool for the post-processing
of spectra produced by Espadons and Narval spectropolarimeters.  Our
inversion tool relies on principal component analysis. It allows
reduction of dimensionality and the definition of a specific metric
for the search of nearest neighbours between an observed spectrum and
a set of synthetic spectra. Effective temperature, surface gravity,
total metallicity and the projected rotational velocity are derived.
Our first tests, done from the sole information coming from a spectral
band very similar to the one the RVS spectrometer will observe from
the Gaia space observatory, and with spectra mainly taken from FGK
stars are very promising. We also tested our method with a few targets
beyond this domain of the H--R diagram.

\end{abstract}



\section{Introduction}

This study is concerned with the inversion of fundamental stellar
parameters from the analysis of high-resolution Echelle spectra.
Hereafter, we shall focus indeed on data collected since 2006 with the
Narval spectropolarimeter mounted at the 2-m aperture
\emph{T\'elescope Bernard Lyot} (TBL) telescope located at the summit
of the \emph{Pic du Midi de Bigorre} (France). We investigate, in
particular, the capabilities of the principal component analysis
(hereafter PCA) for setting-up a fast and reliable tool for the
inversion of stellar fundamental parameters from these high-resolution
spectra.

The inversion of stellar fundamental parameters for each target that
was observed with both Narval and Espadons spectropolarimeters
constitutes an essential step towards: \emph{(i)} the further
post-processing of the data like e.g., the extraction of polarimetric
signals (see e.g., Paletou 2012) but also \emph{(ii)} the exploration,
or data mining, of the full set of data accumulated over the last 8
years now. In Section 2, we briefly describe the actual content of
such a database.

PCA have been used for stellar spectral classification since Deeming
(1964). It has been in use since, and more recently for the purpose of
the \emph{inversion} of stellar fundamental parameters from the
analysis of spectra of various resolutions. It is however most often
used \emph{together} with artificial neural networks (see e.g.,
Bailer-Jones 2000; Re Fiorentin et al. 2007). PCA is used there for
reducing the dimensionality of the spectra before attacking a
multi-layer perceptron which, in turns, allows to link input data to
stellar parameters.

Our usage of PCA for such an inversion process is strongly influenced
by the one routinely made in \emph{solar} spectropolarimetry during
the last decade after the pionneering work of Rees et al. (2000). Very
briefly, the reduction of dimensionality allowed by PCA is directly
used for building a specific metric from which a nearest neighbour(s)
search is done between an observed data set and a learning
database. The latter is made of synthetic data generated from input
parameters properly covering the {\it a priori} range of physical
parameters expected to be deduced from the observations themselves. A
quite similar use of PCA was also presented for classification and
redshift of galaxies estimation by Cabanac et al. (2002). Fundamental
elements of our method and its basic capabilities are exposed in
Sections 3 and 4. One of its originality relies on the simultaneous
inversion of the effective temperature $T_{\rm eff}$, the surface
gravity log$g$, the full metallicity [M/H] and the projected
rotational velocity $vsini$.

In this study we restrict ourselves, on purpose, to the spectral
domain that will be observed by the RVS spectrometer on-board the Gaia
spacecraft (Katz et al. 2004). The RVS will finally operate in a
spectral domain of the order of 847--871 nm (F. Th\'evenin, private
communication). However, for our study we shall use the nominal
spectral domain mentioned up to 2012 in the litterature and which
covers the spectral domain 847--874 nm, at the vicinity of the
Ca\,{\sc ii} infrared triplet. The pertinence of this choice for the
further characterization of the observed stars was discussed by Munari
(1999). This allows us to anticipate reasonable stellar parameter
inversions for stars from B8 to M8, a very large range
of spectral types similar to the actual content of our database.

The conditioning of observed spectra prior to their ingestion into our
inversion tool is detailed in Section 5, and first tests made with
solar spectra observed by reflection over the surface of the Moon are
discussed in Section 6.  Additional tests are also presented and
discussed in Section 7, mainly for FGK-dwarf stars for which
fundamental parameters are already available from the so-called Spocs
catalogue (Valenti \& Fisher 2005). We then briefly discuss
preliminary tests using spectra from giant stars as well as hotter and
cooler than FGK stars.


\section{The source of data}

We mainly used Narval data available from the \emph{public} database
TBLegacy\footnote{http://tblegacy.bagn.obs-mip.fr/ -- note that one
  should rather use the more comprehensive {\tt polarbase.irap.omp.eu} now}. Narval is a
state-of-the-art spectropolarimeter operating in the 0.38-1 $\mu$m
spectral domain, with a spectral resolution of 65\,000 in its
polarimetric mode. It is an improved copy, adapted to the 2-m TBL
telescope, of the Espadons spectropolarimeter, in operations since
2004 at the 3.6-m aperture CFHT telescope.

The TBLegacy database is operational since 2007. It is at the present
time the largest on-line archive of high-resolution polarization
spectra. It hosts data that were taken at the 2-m TBL telescope since
December 2006. So far, more than 70\,000 spectra have been made
available, for more than 370 distinct targets all over the
Hertzsprung-Russell diagram. More than 13\,000 \emph{polarized}
spectra are also available, mostly for \emph{circular} polarization.
Linear polarization data are very seldom still and amount to a few
hundreds spectra, but they are equally available.

\begin{figure}
  \includegraphics[width=9.5cm,angle=0]{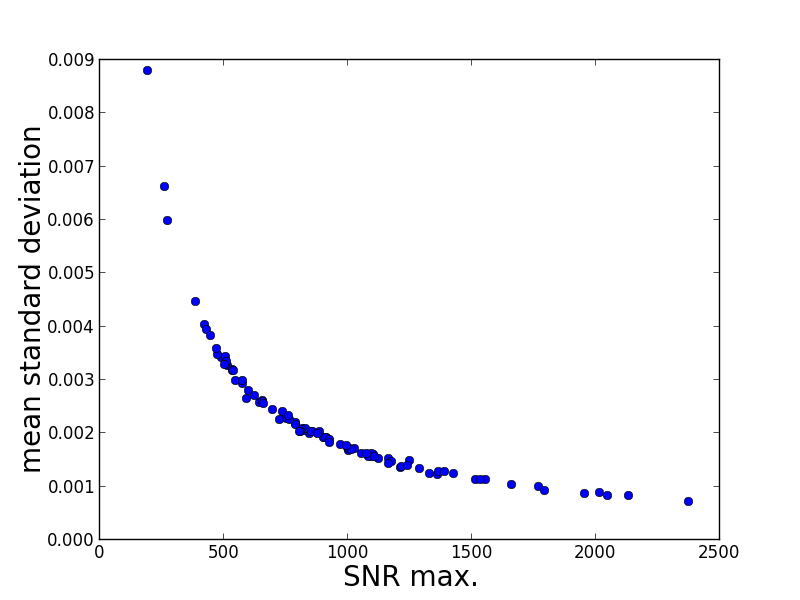}
  \caption{Typical domain of variation of the noise level associated
    with the Narval spectra we shall process.  The standard deviation
    of noise per pixel, for the wavelength range around the infrared
    triplet of Ca\,{\sc ii} is displayed here vs. the maximum signal to
    noise ratio of the full spectra.}
  \label{Fig1}
\end{figure}

At the present time, the TBLegacy database provides no more than
Stokes $I$ or $V /I_c$ spectra calibrated in wavelength. Stokes $I$
data are either normalized to the local continuum or not. We have
however plans: \emph{(i)} to extend it to Espadons data from the CFHT
telescope and \emph{(ii)} to propose higher-level data, such as
pseudo-profiles resulting from line addition and/or least-squares
deconvolution (see e.g., Paletou 2012), activity indexes as well as
stellar fundamentals parameters. The latter's knowledge, besides being
obviously interesting by itself, is also indispensable to any accurate
further post-processing of these high-resolution spectra. These
spectra are also generally bearing high signal-to-noise ratios, as can
be seen in Fig.\,(\ref{Fig1}). Indeed Stokes $I$ spectra we have been
using result from the combination of 4 successive exposures, each of
them carrying 2 spectra of orthogonal polarities generated by a Savart
plate-type analyser. This procedure of double so-called
``beam-exchange'' measurement is indeed meant for the purpose of
extracting (very) weak polarization signals (see e.g., Semel et
al. 1993).

\section{PCA inversion}

Our PCA inversion tool is strongly inspired by magnetic and velocity
field inversion tools which have been developed during the last decade
to complement solar spectropolarimetry (see e.g., Rees et
al. 2000). Improvements of this method have been recently exposed by
Casini et al. (2013) for instance.  Hereafter we describe its main characteristics,
in the context of our study.

\subsection{The training database}

Let us call \{$S_{i} (\lambda)$\} our training database of
\emph{synthetic} stellar spectra. Each of these spectra is
characterised by a limited number of physical parameters which serve
as input parameters for the numerical code producing them. Usually,
for so-called \emph{standard} stellar models, the minimal set of
parameters is the effective temperature of the star, $T_{\rm eff}$,
its surface gravity, log$g$, its \emph{total} metallicity [M/H] (even
though the specific iron abundance [Fe/H] is also frequently used) and
a so-called microturbulent velocity $\xi$ which is an artificial
contributor to line widths, in addition to thermal (or Doppler)
broadening.

Because a significant part of our database is constituted of
moderately, say $vsini > 10 \, {\rm km} \, {\rm s}^{-1}$ to fast
rotators, we adopted the collection of spectra already computed and
made available by Munari et al. (2005). They used Castelli \& Kurucz'
Atlas code and they considered parameters spanning ranges from
3\,500\,K to 47\,500 K for $T_{\rm eff}$, 0. to 5. for log$g$, -2.5 to
0.5 for [M/H], two distinct values, 0. and 0.4 for [$\alpha$/Fe],
microturbulent velocities $\xi$ from 1 to 4 km\,${\rm s}^{-1}$ and
projected rotational velocities v$sini$ ranging from 0 to 500
km\,${\rm s}^{-1}$.

Our training database contains about 34\,757 synthetic spectra after
we limited ourselves to those spectra for which [$\alpha$/Fe]=0, which
may affect our determinations of [M/H] for some objects, and $\xi=2
\,{\rm km}\,{\rm s}^{-1}$ which choice may also not be optimal for all
of our observations. Also, only the so-called ``new ODF'' (where ODF
stands for opacity distribution function; see Castelli \& Kurucz 2003)
spectra were selected for $T_{\rm eff} < 10\,000$ K. Hereafter we
shall focus on the inversion of the set of four parameters \{$T_{\rm
  eff}$; log$g$; [M/H]; $vsini$\} for each of our observed spectra.

It is also important to note that we used Munari et al. (2005) spectra
computed for a resolving power of ${{\cal R} = 20\,000}$ i.e., about a
factor of 3.25 \emph{less} than the one of the observed spectra we
want to process. We shall come back to this point in Sections 5 and 9.

\subsection{Reduction of dimensionality}

Each spectra from Espadons and Narval spectropolarimeters provides of
the order of 250\,000 flux measurements vs. wavelength across a
spectral range spanning from 0.38 to 1 $\mu$m typically. Hereafter we
consider only spectra obtained in the polarimetric mode at a
resolvance ${{\cal R} = 65\,000}$.  Indeed, one of our main objective
is that stellar parameters derived from Stokes $I$ spectra can be
directly used for the further post-processing of the multi-line
\emph{polarized} spectra which comes together (see e.g., Paletou 2012 and
references therein).

For this assessment study, we put ourselves on purpose in a tough
situation by restricting the spectral domain from which we shall
invert stellar parameters to the (initial) one of the RVS instrument
of the Gaia space mission, that is for wavelengths ranging from about
847 to 874 nm (Katz et al. 2004). Arguments in favor of the use of
this very spectral domain can also be found in Munari
(1999). Considering this, the matrix $\vec{S}$ representing our
training database turns to be $N_{\rm models}=34\,757$ by
$N_{\lambda}=1\,569$.

Next we compute the eigenvectors {$\vec{e}_k (\lambda)$} of the
variance-covariance matrix defined as

\begin{equation}
\vec{C} = {(\vec{S}- \bar{S})^{T}} \cdot {(\vec{S}- \bar{S})} \, ,
\end{equation} 
where $\bar{S}$ is the mean of $\vec{S}$ along the $N_{\rm
  models}$-axis.  Therefore $\vec{C}$ is a $N_{\lambda} \times
N_{\lambda}$ matrix.  In the framework of principal component
analysis, reduction of dimensionality is achieved by representing the
original data by a limited set of projection coefficients

\begin{equation}
{p}_{jk}= ({S}_j - \bar{S}) \cdot \vec{e}_k   \, ,
\end{equation} 
with $k_{\rm max} \ll N_{\lambda}$. In what follows for the
processing of all observed spectra, we shall adopt $k_{\rm max}=14$.

\begin{figure}
  \includegraphics[width=9.5cm,angle=0]{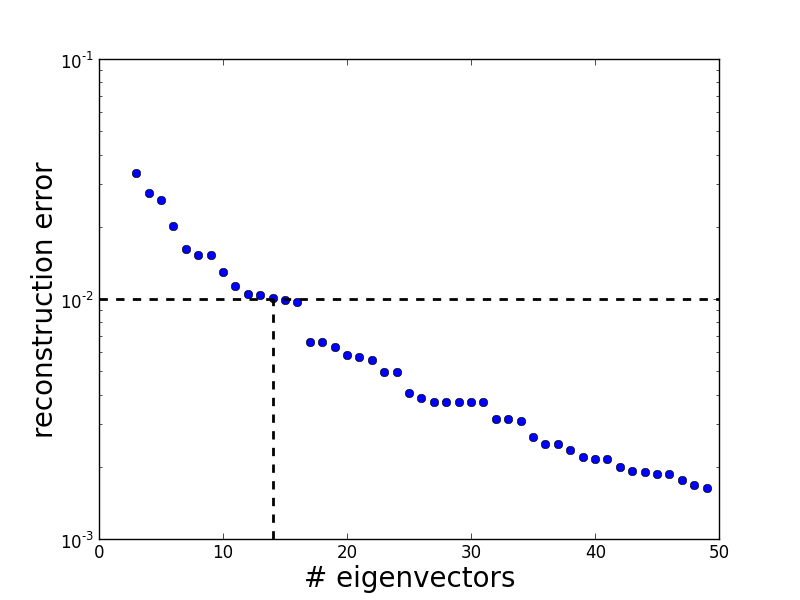}
  \caption{Reconstruction error as a function of the number of
    eigenvectors used for the computation of the admixture
    coefficients $p$.}
  \label{Fig2}
\end{figure}

The most frequent argument supporting the choice of $k_{\rm max}$
relies on the accuracy achieved for the reconstruction of the original
set of $S_i$'s from a limited set of eigenvectors (see e.g., Rees et
al. 2000 or Ram\'{i}rez V\'elez et al. 2010).  In the present case, we
display in Fig.\,(\ref{Fig2}) the maximum reconstruction error

\begin{equation}
  E(k_{\rm max}) = \max\limits_{j} \left( {  { \mid\bar{S} + \sum_{k=1}^{k _{\rm max}} { {p}_{jk} \vec{e}_k } -
        {S}_j \mid } \over { S_j }  }  \right) \, ,
\end{equation} 
as a function of the maximum number of eigenvectors considered for the
computation of the so-called \emph{admixture} coefficients $p$. It is
noticeable that this reconstruction error is better than 1\% for
$k_{\rm max} \ge 14$. However the potentiel effect of noise present in
the observations we want to process should also be taken into account
(see e.g., Socas-Navarro et al. 2001). We shall discuss this specific
point in the next Section.

Practically, we also found convenient to use, at every $k$ order,
${p}_{jk}$'s both centered to their average ${\bar{p}}^{(k)}$ and
normalized to their standard deviation $\sigma_{p}^{(k)}$. It is a
common practice in the field of pattern recognition, since it
guarantees that those coefficients from which we shall define the
specific metric used for the nearest neighbour(s) search will have
comparable effects (on the comparison between spectra).

\subsection{Nearest neighbour(s) search}

The above described reduction of dimensionality allows one to perform
a fast and reliable inversion of observed spectra, once the latter
have been: \emph{(i)} corrected for the wavelength shift vs. synthetic
spectra because of the radial velocity of the target, \emph{(ii)}
continuum-renormalized as accurately as possible, \emph{(iii)}
degraded in spectral resolution to be comparable to the ${{\cal R} =
  20\,000}$ of the synthetic spectra and, finally \emph{(iv)}
resampled in wavelength as the collection of synthetic spectra. We
shall come back to these various stages in the next section. However
once these tasks have been achieved, the inversion process is the
following.

Let $O(\lambda)$ the observed spectrum made comparable to synthetic
ones. We now compute the reduced set of projection coefficients

\begin{equation}
{\varrho}_{k}= { { ({O} - \bar{S}) \cdot \vec{e}_{k} - {\bar{p}}^{(k)}
  } \over {\sigma_{p}^{(k)} } }  \, ,
\end{equation} 
where ${\bar{p}}^{(k)}$ and $\sigma_{p}^{(k)} $ are, respectively, the
mean of the standard deviation of the $ {p}_{jk}$'s, for $k=1, \ldots
, k_{\rm max}$. The nearest neighbour search is therefore done by
seeking the minimum of the squared \emph{Euclidian} distance

\begin{equation}
d^{(O)}_{j} = \sum_{k=1}^{k_{\rm max}}  \left( {\varrho}_{k} - {p}_{jk} \right)^{2} \, ,
\end{equation}
where $j$ spans the number, or a limited number if any {\it a priori} is
known about the target, of distinct synthetic spectra in the training
database.  In practice, we do not limit ourselves to the nearest
neighbour search, although it already provides a relevant set of
stellar parameters. Because PCA-distances between several neighbours
may be of the same order, we adopted a procedure which consists in
considering \emph{all} neighbours in a domain

\begin{equation}
{\rm min} \left ({d}^{(O)}_{j} \right) \leq  {d}^{(O)}_{j} \leq
1.2\times {\rm min} \left( {d}^{(O)}_{j} \right)
\, ,
\label{eq:NNB}
\end{equation} 
and derive stellar parameters as the (simple) mean of each set of
parameters \{$T_{\rm eff}$; log$g$; [M/H]; $vsini$\} characterising
this set of nearest neighbours (A. L\'opez Ariste, private
communication). We did not notice significant changes in the results
either for a smaller range of PCA-distances or when adopting e.g.,
distance-weighted mean parameters.

\section{The effect of noise on the inversion process}

Before processing any real observed spectra with our PCA-based
inversion tool, we need to investigate on the potential effect of
noise. To do so, we have ingested in our inversion tool multiple
realisations of the content of the very learning database
$S_{i}(\lambda)$ affected by controlled (gaussian) white noise-levels,
and we estimated how it affects the inversion process by
systematically comparing (known) input and inverted parameters.

\begin{table}
  \caption{Standard deviation of the absolute differences between input and
    inverted parameters for noisy spectra vs. a (gaussian) white-noise level
    characterized by a $\sigma_{\rm noise}$ per pixel standard
    deviation.}
\label{table:1}
\centering
\begin{tabular}{ccccc}
  \hline\hline
  $\sigma_{\rm noise}$& ${\Delta T_{\rm eff}}$ [K] &
  ${\Delta {\rm log}g}$ & {$\Delta$ [M/H]} &
  ${\Delta v sini} \, [{\rm km}\,{\rm s}^{-1}] $ \\
  \hline
  0.05 & 200 & 0.52 & 0.21 & 15 \\
  0.02 & 101 & 0.27 & 0.10 & 5.5 \\
  0.01 & 46 & 0.12 & 0.04 & 2.6 \\
  0.005 & 10 & 0.04 & 0.01 & 0.5 \\
  \hline
\end{tabular}
\end{table}

A first test was to check upon the choice of $k_{\rm max} = 14$
besides the argument already mentioned concerning the maximum
reconstruction error. In fact, for noise levels characterized by a
standard deviation per pixel better or equal to $\sigma_{\rm
  noise}=0.01$ -- see also Fig.\,(1) -- internal errors are quite
similar and minimal in the $k_{\rm max} \sim 12-15$ range. Typical
values for each stellar parameter are given in Table 1. We could also
check that these values start to significantly increase for values
$k_{\rm max} \ge 16$, which confirmed to us the optimal choice of
$k_{\rm max} =14$ we made for the remainder of this study.

Results of the numerical experiments we summarized in Table 1 also
describe the effect of signal to noise ratio on the quality of our
inversion process. We did not notice any significant bias so that our
values can also be compared to ``mean maximum errors'' used in other
studies (e.g., Recio-Blanco et al. 2006 or Katz et al. 1998). It is
also worth mentioning that, for increasing signal-to-noise ratio
(i.e., decreasing value of $\sigma_{\rm noise}$) although we computed
standard deviations on the distribution of the absolute errors between
input and inverted parameters, these distributions of errors showed to
be strongly skewed towards 0 and departing from gaussian. Finally,
internal errors are about a factor of 2 better, if one limits the
domain of work to $4000 \le T_{\rm eff}\le 7000$ K, log$g \ge 3.0$ dex
and [M/H]$\ge -1.0$ dex according to the domain of parameters relevant
for the Spocs--Narval targets we analysed in this study.

\section{Conditioning of the observed spectra}

The first and obvious task to perform on \emph{observed} spectra is to
correct for their wavelength shift vs. synthetic spectra computed at
radial velocity $V_{\rm rad}=0$. The radial velocity of the target at
the time of the observation is deduced from the centroid, in a
velocity space, of the pseudo-profile resulting from the ``addition''
(see e.g., Paletou 2012) of the three spectral lines of the Ca\,{\sc
  ii} infrared triplet whose rest wavelengths are, respectively,
849.802, 854.209 and 866.214 nm. Once $V_{\rm rad}$ is known the
observed profile is set on a new wavelength grid, at rest. We could
check, using solar spectra (see \S 6.) that $V_{\rm rad}$ should be
known to an accuracy of the order of $\delta v/3$ with $\delta
v=c/{\cal R}$ i.e., in our case about 1.5 ${\rm km}\,{\rm
  s}^{-1}$. Beyond this value, estimates of $vsini$ and, to a minor
extent $T_{\rm eff}$, start to be significantly affected by the
misalignement of the observed spectral lines with those of the
synthetic spectra of the training database.

A second step consists in degrading the resolution of the initial
spectra to the one of the synthetic spectra computed for ${{\cal R} =
  20\,000}$. This is done by convolving the initial observed profile
by a Gaussian profile of adequate width. Then we resample the
wavelength grid down to the one common to all synthetic spectra, and
we interpolate the original spectra onto the new wavelength grid.

Finally, we have to correct for unproper normalization of the Stokes
$I$ flux to the local continuum. This issue has been very well
discussed in Gazzano et al. (2010) as well as in Kordopatis et
al. (2011), and we adopted their iterative procedure. For the FGK
stars spectra we have been mainly dealing with hereafter, we report no
more than 3.5\% initial relative error in position for the continuum
level, while in most case this does not exceed 2\%. The iterative
correction procedure allows to renormalize our observed spectra at
$10^{-4}$, using successive estimates of the nearest neighbour
synthetic spectra as a reference. This procedure is clearly required,
and we also report on possible errors of the order of $\Delta T_{\rm
  eff}\sim 150$ K, $\Delta {\rm log}g \sim 0.1$ dex and $\Delta {\rm
  [M/H]} \sim 0.1$ dex due to unproper normalization of the continuum.

\section{First tests with observed solar spectra}

\begin{figure}
  \includegraphics[width=9.5cm,angle=0]{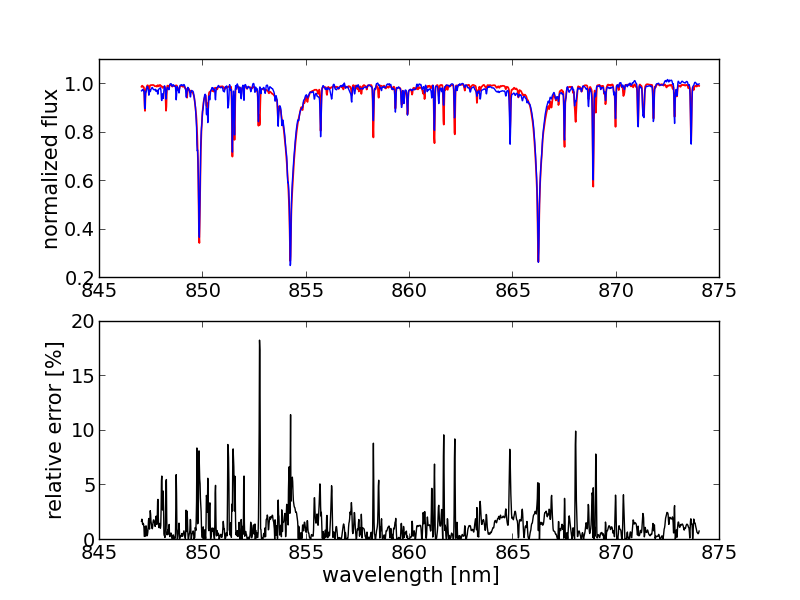}
  \caption{The top figure displays normalized flux from the
    observation of the Sun by reflection over the Moon made at the TBL
    with the Narval spectropolarimeter (blue) and the nearest
    PCA-distance synthetic spectrum (red). The latter corresponds to stellar
    parameters $T_{\rm eff}$=6000 K, log$g$=4.5, [M/H]=0 and $vsini$=5
    ${\rm km}\,{\rm s}^{-1}$.  The bottom figure displays the relative
    error vs. wavelength between the two spectra.}
  \label{Fig3}
\end{figure}

First tests of our inversion method with real data were performed
using solar spectra observed by the 2-m aperture TBL telescope by
reflection over the surface of the Moon in March and June 2012.

The synthetic spectrum having the minimal PCA-distance with the
observed spectrum have model parameters $T_{\rm eff}$=6000 K,
log$g$=4.5, [M/H]=0 and $vsini$=5 ${\rm km}\,{\rm s}^{-1}$. It is
displayed against one of our observed spectra in Fig.\,(\ref{Fig3}).
These values slightly overestimate ``canonical'' values of $T_{\rm
  eff}$=5780 K, log$g$=4.4 and $vsini$ about 2 ${\rm km}\,{\rm
  s}^{-1}$ (see e.g., Pavlenko et al. 2012). However, considering the
``bulk'' of nearest neighbours in the range defined by inequalities
(\ref{eq:NNB}), we derive more accurate parameters such as $T_{\rm
  eff}$=5772 K, log$g$=4.26, [M/H]=0 and $vsini$=2.2 ${\rm km}\,{\rm
  s}^{-1}$. Typically of the order of 20 neighbour-models are
identified with our procedure and for Narval solar spectra.

The relevance of the PCA-distance on which rely our inversion process
can also be verified by the examination of the characteristics of the
set of nearest neighbours we could identify. In the solar spectrum
case, we find: 9 models bearing $T_{\rm eff}$=5750 K, 6 models with
$T_{\rm eff}$=5500 K, as many at 6000 K and one at 6250 K. For the
surface gravity we find: 9 with log$g$=4.0, 6 with 3.5, as many at 4.5
and just one at 5.0. Finally, concerning the projected rotational
velocity, we find: 8 models at 0, 7 at 2 ${\rm km}\,{\rm s}^{-1}$ and
as many at 5 ${\rm km}\,{\rm s}^{-1}$. Note finally that \emph{all}
neighbours have a solar metallicity [M/H]=0. From this point of view,
it seems that we are doing better than Kordopatis et al. (2011) who
state that they do not recover a metallicity better than -0.1 dex for
a solar spectra using their pipeline combining the Matisse
(projection) method (Recio-Blanco et al. 2006) and a $k$d-tree
classification scheme.

\begin{table*}
  \caption{Summary of the comparison between our inverted
    parameters and these provided: (1) by the Spocs catalogue and (2) in
    the litterature i.e., using Simbad at CDS and/or the Pastel catalogue
    (Soubiran et al. 2010), for the objects
    listed below.}
\label{table:2}
\centering
\begin{tabular}{ccccccccc}
\hline\hline
Object & $T_{\rm eff}^{\rm (Spocs)}$ &$T_{\rm eff}^{\rm (PCA)}$ &
log$g^{\rm (Spocs)}$ & log$g^{\rm (PCA)}$ & [M/H]$^{\rm (Spocs)}$ &
[M/H]$^{\rm (PCA)}$ & ${vsini}^{\rm (Spocs)}$ & ${vsini}^{\rm (PCA)}$ \\
\hline
HD 377 & 5873 & 5644 & 4.3 & 4.70 & 0.1 & 0.0 & 14.6 & 15.2 \\
HD 120136 & 6387 & 6366 & 4.3 & 4.68 & 0.2 & 0.2 & 15.0 & 19.3 \\
HD 12328 & 4919 & 4761 & 3.7 & 3.63 & -0.1 & -0.03 & 1.9 & 2.6 \\
\hline\hline
Object & $T_{\rm eff}^{\rm (Litt.)}$ &$T_{\rm eff}^{\rm (PCA)}$ &
log$g^{\rm (Litt.)}$ & log$g^{\rm (PCA)}$ & [M/H]$^{\rm (Litt.)}$ &
[M/H]$^{\rm (PCA)}$ & ${vsini}^{\rm (Litt.)}$ & ${vsini}^{\rm (PCA)}$ \\
\hline
GL 205 & 3\,730 & 3\,750 & 4.73 & 4.8 & 0.0 & 0.0 & 2.73 & 3. \\
Arcturus & 4\,290 & 4\,250 & 1.7 & 2.0 & -0.5 & -0.5 & 4.2 & 5. \\
Sirius & 9\,830 & 10\,500 & 4.3 & 4.5 & 0.34 & 0.5 & 16. & 20. \\
\hline
\end{tabular}
\end{table*}

\section{Other FGK stars}

Because one of the originality of our inversion tool is in providing
the projected rotational velocity of stars, we also tested its
capability for a couple of FGK stars with significant ($\sim 15-20
\,{\rm km}\,{\rm s}^{-1}$) $vsini$ and belonging the so-called ``Stellar
properties of observed cool stars'' (aka. Spocs) catalogue (Valenti \&
Fischer 2005).

For a first test, we used a spectra from the pre-main sequence star
HD377 and comparison between Spocs parameters and ours are detailed in
Table \ref{table:2}. Unfortunately, according to the Pastel catalogue
(Soubiran et al. 2010) there are almost no alternative values given in
the litterature for this object. Its $vsini$ is well recovered and
other parameters agree reasonably well with Spocs values. Our second
target is $\tau$ Boo (A, also HD120136), a F6IV star for which the
Spocs catalogue provides a $vsini$ of 15 km\,${\rm s}^{-1}$ while our
estimate is about 19 km\,${\rm s}^{-1}$. However a $vsini$ of about 18
km\,${\rm s}^{-1}$ has been recently reported by
Mart\'{\i}nez-Arn\'aiz et al. (2010). Other parameters agree very
well, except for log$g$ that we systematically overestimated for these
two objects.

A first excursion away from the FGK-dwarfs domain consists in
exploring the luminosity class towards the giants domain. The red
giant branch (G5) star HD12328 have been used for that purpose. Our
results are again reliable given also that recent works mention an
effective temperature of 4808 K, in better agreement with ours
(Massarotti et al. 2008), as well as log$g$ value of the order of 3.3
dex and [Fe/H]=-0.04 dex (see also Jones et al. 2011). Outside the
Spocs catalogue, and again for (sub-)giants, we also used data from
the K1.5III star Arcturus. Our 4250 K determination of its effective
temperature is very close, within 30 K, to that of Prugniel et
al. (2011) for instance. We could also derive a surface gravity about
2.0 in agreement with alternative estimates (e.g., Massarotti et
al. 2008) and our [M/H] and $vsini$ values are also in pretty well
identified ranges, as reported in Table 2. Note that further tests
with Pollux (K0III) data also gave excellent results. Concerning these
classes of stars, we are also perfectly aware that additional
synthetic spectra with microturbulent velocities not restricted to 2
km\,${\rm s}^{-1}$ as we did here, will have to be next
considered. Note also that for (sub-)giant stars, models computed in
(1D) \emph{spherical} geometry such as MARCS ones (Gustafsson et
al. 2008) may also provide a more realistic description of the
observed stellar fluxes and therefore improve the determination of the
inverted stellar parameters.

\section{Beyond the FGK domain}

Our database of spectropolarimetric data already covers pretty well
the H-R diagram. We are therefore interested in spanning most of it,
and basically with the same inversion tool. Selected targets for these further
tests are indicated, together with the parameters we derived from
their spectra, in Table 2.

For cooler stars, we picked spectra of GL 205, a M1.5V dwarf. Our
values of $T_{\rm eff}$, log$g$ and [M/H] agree very well with those
of Prugniel et al. (2011), and our $vsini$ determination agrees well
with the one of Houdebine (2010). There is also a reasonable agreement
with values recently derived by Neves et al. (2013).

\begin{figure}
  \includegraphics[width=9.5cm,angle=0]{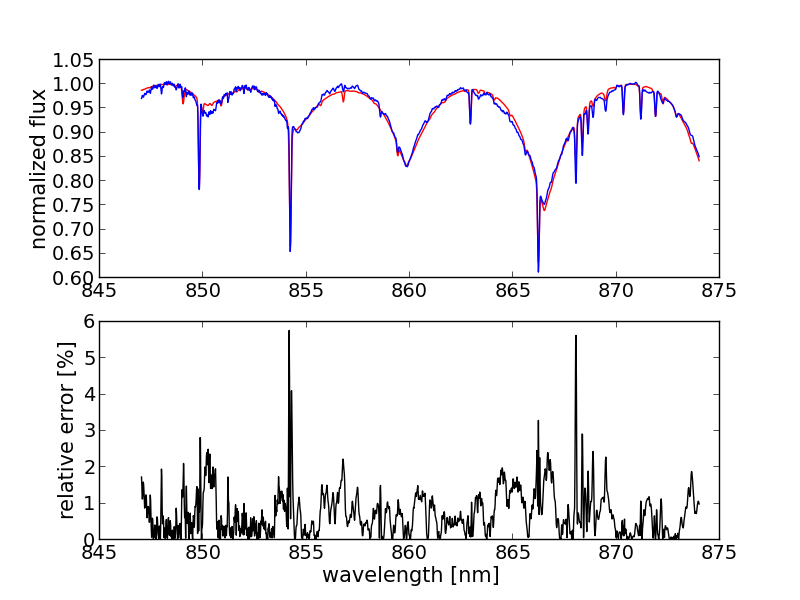}
  \caption{Same as Fig. (3) but for Sirius A.}
  \label{Fig4}
\end{figure}

Finally, we also tested our method with spectra of the A1V star Sirius
(A). Our inversion of the effective temperature gives a value a bit
larger than the 9870 K more recently reported (Hill \& Landstreet
1993). [M/H] is also overestimated, but log$g$ and $vsini$ are quite
well determined. We are also aware that, so far, we restricted
ourselves to models such that [$\alpha$/Fe]=0 which may not apply to
such a star. Possible enhancement of $\alpha$-elements will be
considered in a next version of our inversion tool, when performing
spectral analysis at a larger scale than the one presented
here. However, besides this current restriction, the agreement between
the observed spectrum and the nearest neighbour synthetic spectrum is
already quite satisfactory, as shown in Fig.\,(\ref{Fig4}).

\section{Discussion}

Concerning the determination of $vsini$, other methods of evaluation
exist. However, to the best of our knowledge, they always require a
template spectrum or at least a list of spectral line {\it a priori}
expected in the spectra, as auxilliary and ``support'' data (see e.g.,
D\'{\i}az et al. 2012 and references therein). Our final pipeline will
therefore implement an additional module providing, once a preliminary
stellar parameter inversion will be available from our inversion tool,
an alternative and complementary $vsini$ evaluation.  Note also that,
with our PCA method, we are mostly interested in the ``intermediate''
$vsini$ regimes, say between 10 and 100 km\,${\rm s}^{-1}$. Indeed, for
slow rotators it is obvious that rotational broadening becomes of the
order of other sources of broadening (e.g., instrumental or turbulent)
so that a more detailed line profile analysis may be required. On the
other hand, (very) fast rotators are affected by macroscopic effects such as
gravity darkening (see e.g., Espinosa Lara \& Rieutord 2011) which are
not accounted for, so far, by standard atmospheric (radiative)
modelling.

Another source of potential improvement relies on the content of our
learning database. ATLAS standard models may not be the best choice
for cool stars or for metal-poor stars for instance. It is also
well-known that non-LTE effects may take place in the formation of the
infrared triplet of Ca\,{\sc ii}, which affects the spectral lines
central depressions. This issue was for instance mentioned in
Kordopatis et al. (2011). More detailed elements, from the point of
view of radiative modelling, are also discusses in Merle et
al. (2011). Finally, we shall need synthetic spectra computed for a
resolvance comparable to the one of our observations and, ideally,
including the effects of rotational broadening. The database of
synthetic spectra, Pollux (Palacios et al. 2010) already provides a
partial answer to our needs, and we believe that its development will
take into account such needs.

\section{Conclusion}

We have experimented a fast and reliable PCA-based numerical method
for the inversion of stellar fundamental parameters $T_{\rm eff}$,
log$g$ and [M/H], as well as the projected rotational velocity $vsini$,
from high-spectral resolution Echelle spectra taken from the Narval
spectropolarimeter in operations at the 2-m TBL telescope. 

Our method is fast and easy to implement. First tests, mainly made for
FGK-stars spectra show fairly good agreement with reference values
published by Valenti \& Fischer (2005). Preliminary tests also suggest
that the same method is applicable to FGK (sub-)giants, as well as to
cooler M stars or hotter stars up to spectral type A.  We used it, so
far, at the vicinity of the infrared triplet of Ca\,{\sc ii} and
without any help from additional (e.g., photometric) information,
which is particularly challenging. However we can easily, either
extend the spectral domain used by our inversion method, or combine
analyses from several distinct spectral domains, in order to constrain
further and refine our stellar parameter determination.

It is finally important to realize from the present study that PCA
allows for a reduction of dimensionality of the order of 100 ($\sim
N_{\lambda}/k_{\rm max}$) which is of great potential interest for the
post-processing of high-resolution spectra covering a very large
bandwidth like the ones from both Espadons and Narval
spectropolarimeters.

\section*{Acknowledgments}
  We are grateful to Dr. Tomaz Zwitter (University of Ljubljana,
  Slovenia) who made available the synthetic spectra used in the work.
  This research has made use of the VizieR catalogue access tool, CDS,
  Strasbourg, France. The original description of the VizieR service
  was published in A\&AS 143, 23. This research has made use of the
  SIMBAD database, operated at CDS, Strasbourg, France. Narval data
  were provided by the OV-GSO datacenter operated by CNRS/INSU and the
  \emph{Universit\'e Paul Sabatier, Toulouse}-OMP (Tarbes, France;
  {\tt polarbase.irap.omp.eu}).

\label{lastpage}

\end{document}